\makeatletter
\@ifundefined{@parse@version@dash}{%
\def\@parse@version#1{\@parse@version@0#1}
\def\@parse@version@#1/#2/#3#4#5\@nil{%
\@parse@version@dash#1-#2-#3#4\@nil}
\def\@parse@version@dash#1-#2-#3#4#5\@nil{%
  \if\relax#2\relax\else#1\fi#2#3#4 }
}{}
\makeatother
\documentclass[reprint, amsmath,amssymb, pra]{revtex4-2}

\usepackage{tikz}
\usepackage{color}
\usepackage{ulem}
\usepackage{graphicx}
\usepackage{dcolumn}
\usepackage{bm}
\usepackage[colorlinks, linkcolor=blue, anchorcolor=blue, citecolor=blue, urlcolor=blue]{hyperref}
\usepackage{url}
\definecolor{lime}{HTML}{A6CE39}
\DeclareRobustCommand{\orcidicon}{%
	\begin{tikzpicture}
	\draw[lime, fill=lime] (0,0)
	circle [radius=0.16]
	node[white] {{\fontfamily{qag}\selectfont \tiny ID}};
	\draw[white, fill=white] (-0.0625,0.095)
	circle [radius=0.007];
	\end{tikzpicture}
	\hspace{-2mm}
}

\foreach \x in {A, ..., Z}{%
	\expandafter\xdef\csname orcid\x\endcsname{\noexpand\href{https://orcid.org/\csname orcidauthor\x\endcsname}{\noexpand\orcidicon}}
}

\begin{document}

\title{Quantum gyroscopes based on double-mode surface-acoustic-wave cavities}
\author{Yuting Zhu\orcidA{}$^{1,2,3}$}
\author{Shibei Xue\orcidB{}$^{1,2,3}$}
\email{shbxue@sjtu.edu.cn}
\author{Fangfang Ju$^{5}$}
\author{Haidong Yuan\orcidC{}$^{4}$}

\affiliation{$^{1}$Department of Automation, Shanghai Jiao Tong University, Shanghai 200240, P. R. China}
\affiliation{$^{2}$Key Laboratory of System Control and Information Processing, Ministry of Education of China, Shanghai 200240, P. R. China}
\affiliation{$^{3}$Shanghai Engineering Research Center of Intelligent Control and Management, Shanghai 200240, P. R. China}
\affiliation{$^{4}$Department of Mechanical and Automation Engineering, The Chinese University of Hong Kong, Shatin, Hong Kong SAR, P. R. China}
\affiliation{$^{5}$School of Physics and Electronics, Hunan Normal University, Changsha 410081, P. R. China}
\date{\today}

\begin{abstract}
Recent progress shows that a surface-acoustic-wave (SAW) cavity can not only induce quantum acoustic dynamics but also can form optomechanical-like systems. Its operating frequencies in the microwave band make it resistant to the thermal noise of surrounding environments, while its radiation-pressure couplings make it susceptible to weak forces. Based on these advantages, we propose a gyroscope comprising coupled microwave-SAW cavities. In this paper, we systematically consider the three indices including range, signal-to-noise ratio, and sensitivity, which are the most important to gyroscopes but only partially considered in existing works. Additionally, we establish the fundamental limits of sensitivity when the quantum input is in the vacuum state and the squeezed vacuum state. We find that squeezing improves sensitivity and can surpass the standard quantum limit. However, this improvement can only reach up to $\sqrt{2}/2$ even as the squeezed parameter approaches infinity, which is rarely noted in recent works. Finally, we also offer analytical constraints for cooperativity and squeezed parameters. These constraints can be utilized to design gyroscopes based on coupled cavities in experiments.
\end{abstract}
\maketitle
\section{Introduction\label{Sec1}}


Gyroscopes have made significant contributions to humankind as sensors for measuring angular velocity since they were first proposed by Foucault in 1852. However, measuring an extremely weak angular velocity, especially one much smaller than the Earth's rotation, poses great challenges for classical gyroscopes such as rigid rotator gyroscopes \cite{Barr1961, Craig1972}, mircoelectromagnetic gyroscopes \cite{Acar2008, Armenise2010}, and Sagnac laser gyroscopes \cite{Chow1985}. These classical gyroscopes struggle to meet the sensitivity and scalability requirements of modern gyroscopes. This directs scientists' attention to quantum gyroscopes, such as atom interferometer gyroscopes \cite{Riehle1991, Gustavson1997, Dowling1998, Gustavson2000, Durfee2006, Canuel2006, Stockton2011,Berg2015, Dutta2016, Savoie2018, Yao2021, Moan2020, Wu2020, Guo2021}, nuclear magnetic resonance gyroscopes \cite{Woodman1987, Kornack2005, Meyer2014, Zhang2020} and optomechanical gyroscopes \cite{Davuluri2016, Davuluri2017, Li2018, Li2018B1, Li2018C1}, as quantum devices are more susceptible to interacted perturbations \cite{Degen2017, Lau2018, Pirandola2018, Pezze2018, McDonald2020, Rudolph2022}.

Among the above quantum gyroscopes, the optomechanical gyroscope stands out for not requiring the construction of magneto-optical trapping \cite{Gustavson1997, Gustavson2000, Canuel2006, Stockton2011, Berg2015, Dutta2016, Savoie2018, Yao2021} or vapor chamber \cite{Woodman1987, Kornack2005, Meyer2014, Zhang2020} to trap ions/atoms. This feature makes it highly suitable for on-chip integration compared to the other types of gyroscopes. Furthermore, optomechanical systems benefit from their susceptibility to weak forces because of their unique radiation-pressure coupling, which enables them to be almost ideal devices for designing quantum gyroscopes based on centrifugal forces \cite{Davuluri2016} or Coriolis forces \cite{Davuluri2017, Li2018, Lavrik2019}. However, the susceptibility to weak forces and the low frequency of the mechanical mode (typically in the kilohertz to megahertz range \cite{Aspelmeyer2014}) also make optomechanical systems susceptible to thermal Langevin forces, which restrict further improvements in sensitivity. For instance, a recent study shows that the sensitivity decreases by approximately six orders of magnitude when the temperature increases from 0 K to 300 K \cite{Davuluri2017}. Fortunately, emerging surface-acoustic-wave (SAW) cavities offer a potential solution to overcome this problem. The SAW cavity is a novel type of mechanical oscillator that operates in the microwave band and demonstrates exceptional quantum coherence \cite{Gustafsson2014, Schuetz2015, Chu2017, Manenti2017, Chu2018, Satzinger2018, Ekstrom2019, Andersson2019, Delsing2019, Noguchi2020, Wu2021,  Zhu2022, Zhu2022A,Guo2017, Yin2023}. In addition, SAW cavities can also be utilized for creating optomechanic-like systems with radiation-pressure couplings \cite{Noguchi2020, Wu2021}. These advantages make SAW cavities not only retain the susceptibility to weak forces but also immune to thermal Langevin forces, unlike the mechanical mode of existing optomechanical systems. Therefore, they are superior devices for designing gyroscopes.

In addition to the devices, three crucial indices that need to be taken into account when designing gyroscopes: (i) the range, which determines the interval within which the angular velocity can be detected; (ii) the signal-to-noise ratio (SNR), which determines whether the output signal can be readout; and (iii) the sensitivity, which determines the minimum detectable change of the angular velocity. These three indices are actually interrelated rather than independent with each other because they are constrained by the same parameters within a given system. Also, these indices are related to the noise of the system and are limited by the standard quantum limit \cite{Clerk2010, Aspelmeyer2014}, i.e., the minimum quantum noise allowed by the Heisenberg uncertainty relation. However, in recent proposals \cite{Wu2020, Guo2021, Davuluri2016, Li2018, Li2018B1, Li2018C1}, the authors either only consider a portion of the above indices or fail to take into account the limitation imposed by the standard quantum limit. For instance, in recent atom interferometer gyroscopes \cite{Wu2020,Guo2021}, the authors only consider the influence of quantum input noise but neglect other coherent quantum noise sources such as thermal Langevin noises. As a result, the best sensitivity obtained only applies to ideal cases that are not constrained by standard quantum limits. In recent optomechanical gyroscopes \cite{Davuluri2016, Li2018B1}, the authors only consider the readable condition $\mathrm{SNR}\geq1$, but this condition actually determines the range of the angular velocity rather than sensitivity. In Ref. \cite{Li2018C1}, the authors fail to consider the readable condition $\mathrm{SNR}\geq1$, although they skillfully employ coherent quantum noise cancellation to break the limitation on output noise imposed by the standard quantum limit. In Ref. \cite{Li2018}, the authors use normal-mode splitting to measure angular velocity. However, this method only provides the range of the angular velocity. Therefore, recent proposals for quantum gyroscopes are incomplete.

To the above ends, we propose a quantum gyroscope utilizing SAW cavities and systematically take all three essential indices into account to overcome problems as mentioned above. Also, we discuss the fundamental limits of sensitivity in detail when the input is the vacuum state and squeezed vacuum state, respectively. As a result, the sensitivity is limited by the standard quantum limit when the input is in the vacuum state, and this limit can be surpassed after squeezing. More importantly, squeezing also has a limit to the improvement in sensitivity, and this limit is $\sqrt{2}/2$. However, this crucial result is rarely noted in recent works \cite{Davuluri2016, Davuluri2017, Li2018, Li2018B1, Li2018C1,Wu2020, Guo2021}. Furthermore,  we provide analytical constraints on the cooperativity and the squeezed parameter, which would be beneficial for experiments.

The remainder of this paper is organized as follows: In Sec. \ref{Sec2}, we provide the model of the gyroscope and the equations of motion according to quantum Langevin equation. We then analyze its range, SNR, sensitivity, and standard quantum limit in Sec. \ref{Sec3}.  In Sec. \ref{Sec4}, we provide the corresponding numerical simulations. Finally, we conclude this work in Sec. \ref{Sec5}.

\section{Quantum Gyroscope and its Langevin equations \label{Sec2}}

\begin{figure} [hbt]
\centering
\includegraphics[width=0.48\textwidth]{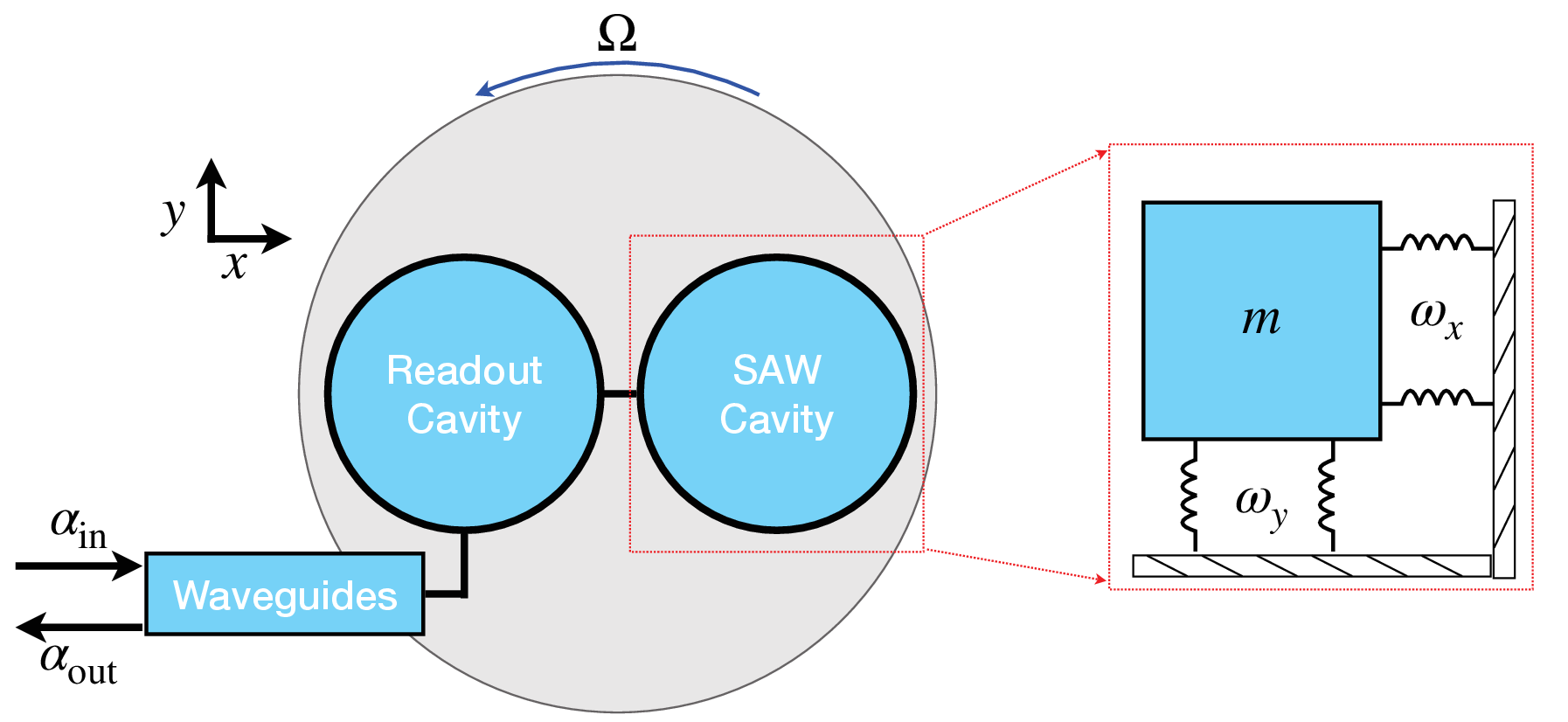}
\caption{(Color online) Schematic of the quantum gyroscope. The system consists of two coupled cavities, where the SAW cavity acts as a mechanical oscillator having the $x$ and $y$ modes, as shown in the red box. These two modes couple to each other via the rotation of the platform with an unknown angular velocity $\Omega$, forming a basic gyroscope scheme in the $x-y$ plane. In addition, the $x$ mode also couples to a readout cavity (microwave band) that is connected to a waveguide. This allows the unknown angular velocity $\Omega$ to be readout through the output $\alpha_{\mathrm{out}}$ using a homodyne detection.}
\label{model}
\end{figure}

The quantum gyroscope being studied is based on previous optomechanical gyroscopes \cite{Davuluri2016, Davuluri2017, Li2018B1, Li2018C1}. It consists of two coupled cavities fixed on a platform that rotates with an unknown angular velocity $\Omega$, as shown in Fig. \ref{model}. The SAW cavity acts as a double-mode mechanical oscillator with an effective mass $m$ and two operating frequencies $\omega_x$ and $\omega_y$, as depicted in the red box in Fig.~\ref{model}. Hereafter, the subscripts $x$ and $y$ are used to label corresponding quantities in the $x$ and $y$ directions, respectively. Also, a readout cavity operating in the microwave band connected to a waveguide is used for homodyne detection. Our aim is to readout the angular velocity $\Omega$ of the platform through the output $\alpha_{\mathrm{out}}$ in the waveguide.

The Hamiltonian of our system in terms of bosonic creation and annihilation operators reads
\begin{equation}
H=H_a+H_m+H_I
\end{equation}
with
\begin{subequations}
\begin{eqnarray}
&&H_a=\hbar\omega_a a^\dag a-i\hbar\sqrt{\kappa}(a^\dag \alpha_{\mathrm{in}}e^{-i\omega_d t}+H.c.)\\
&&H_m=\hbar\omega_x(b_x^\dag b_x+\frac{1}{2})+\hbar\omega_y(b_y^\dag b_y+\frac{1}{2})\label{HM}\nonumber\\
&&+\frac{i\hbar\Omega}{2}[\eta_1(b_x^\dag b_y-b_x b_y^\dag)+\eta_2(b_x^\dag b_y^\dag-b_x b_y)]\\
&&H_I=-i\hbar g_1(b_x a^\dag-b_x^\dag a)+i\hbar g_2(b_x a- b_x^\dag a^\dag).
\end{eqnarray}
\end{subequations}
Here, $H_a$ represents the Hamiltonian of the readout cavity driven by the input $\alpha_{\mathrm{in}}=\alpha+a_{\mathrm{in}}$, where $N_{\mathrm{in}}=|\alpha|^2$ represents the input photon number, and $a_{\mathrm{in}}$ refers to the corresponding quantum  input \cite{Clerk2010}. $H_m$ represents the Hamiltonian of the double-mode SAW cavity with the coefficients $\eta_1=\sqrt\frac{\omega_x}{\omega_y}+\sqrt\frac{\omega_y}{\omega_x}$ and $\eta_2=\sqrt\frac{\omega_x}{\omega_y}-\sqrt\frac{\omega_y}{\omega_x}$ (See Appendix.  \ref{Appendix} for more details). Obviously, the two initially isolated modes become coupled to each other when the platform is rotating, forming a basic gyroscope scheme in the $x-y$ plane. Without loss of generality, the Hamiltonian $H_m$ considered here is a complete form including centrifugal forces which are neglected in the previous work \cite{Li2018B1} (See more details in the Appendix. \ref{Appendix}). $H_I$ denotes the interaction between the readout cavity and the $x$ mode of the SAW cavity. The first term corresponds to a beam-splitter coupling, while the second term represents a down-conversion coupling \cite{Schweer2022}. When the coupling strength $g_1=g_2$, the interaction Hamiltonian $H_I$ reduces to a linearized optomechanical coupling $i\hbar g(a-a^\dag)(b_x+b_x^\dag)$ \cite{Clerk2010, Aspelmeyer2014}. For brevity, we consider that $\omega_x=\omega_y=\omega_b$ and $g_1=g_2=g$ in the following contents.

In a rotating frame with respect to the driving frequency $\omega_d$, the quantum Langevin equations of the system read
\begin{subequations}
\begin{eqnarray}
&&\dot{a}=(i\Delta-\frac{\kappa}{2})a-g(b_x+b_x^\dag)-\sqrt{\kappa}\alpha_{\mathrm{in}},\label{EOMA}\\
&&\dot{b}_x=(-i\omega_b-\frac{\gamma_x}{2})b_x+g(a-a^\dag)+\Omega b_y-\sqrt{\gamma_x}f_x,\label{EOMBX}\nonumber\\
\\
&&\dot{b}_y=(-i\omega_b-\frac{\gamma_y}{2})b_y-\Omega b_x-\sqrt{\gamma_y}f_y\label{EOMBY},
\end{eqnarray}
\end{subequations}
where the detuning between the readout cavity and the driving is $\Delta=\omega_d-\omega_a$, and $\gamma_{x(y)}$ is the decay rate of the $x(y)$ mode due to the thermal noise $f_{x(y)}$. The thermal noises satisfy the following correlations \cite{Wu2021}
\begin{eqnarray}
&&\langle f_{x(y)}(t)\rangle=0,\nonumber\\
&&\langle f^\dag_{x(y)}(t)f_{x(y)}(t')\rangle=n_{\mathrm{th}}\delta(t-t'),\nonumber\\
&&\langle f_{x(y)}(t)f^\dag_{x(y)}(t')\rangle=(n_{\mathrm{th}}+1)\delta(t-t'),
\end{eqnarray}
where $n_{\mathrm{th}}=[\exp~(\hbar\omega_b/k_B T)-1]^{-1}$ is the equilibrium mean thermal phonon number of the SAW cavity. In the state-of-the-art SAW-superconducting-circuits experiments \cite{Noguchi2020, Wu2021}, the temperature of the surrounding environment can reach the milikelvin regime inside a dilution refrigerator, and the frequency of the SAW cavity $\omega_b$ can be enhanced to gigahertz. Therefore, one can let $n_{\mathrm{th}}=0$. We also specialize in the case where the readout cavity is resonantly driven, i.e., $\Delta=0$, which is a common condition for Homodyne detection in experiment. Furthermore, the corresponding input-output relation is given by
\begin{equation}
\alpha_{\mathrm{out}}(t)=\alpha_{\mathrm{in}}(t)+\sqrt{\kappa}a(t).
\end{equation}

The equations of motion \eqref{EOMA}- \eqref{EOMBY} can be easily solved using Fourier transformations $O(\omega)=\int \mathrm{d}t ~O(t)e^{i\omega t}$ and $O^\dag(\omega)=\int\mathrm{d}t~ O^\dag(t)e^{i\omega t}=[O(\omega)]^\dag$. Here, we consider the frequency of the SAW cavity to be much lower than the damping of the readout cavity, i.e., $\omega_b\ll\kappa$. This allows the state of the readout cavity to adiabatically track the motion of the $x$ mode. Therefore, we can approximately use the steady-state solution of Eq. \eqref{EOMA}
\begin{equation}
a(\omega)=-\frac{2g}{\kappa}(b_x(\omega)+b^\dag_x(\omega))-\frac{2}{\sqrt{\kappa}}\alpha_{\mathrm{in}}(\omega)
\end{equation}
to replace its original solution $a(\omega)=[g(b_x(\omega)+b_x^\dag(\omega))+\sqrt{\kappa}\alpha_{\mathrm{in}}(\omega)]/(i\omega-\frac{\kappa}{2})$. In doing so, we then obtain the solution of Eq. \eqref{EOMBX} in the frequency domain
\begin{eqnarray}
b_x(\omega)=\chi_x(\omega-\omega_b)[\sqrt{\gamma_x C_o}(\alpha_{\mathrm{in}}(\omega)-\alpha^\dag_{\mathrm{in}}(\omega))\nonumber\\
+\sqrt{\gamma_x}f_x(\omega)-\Omega \chi_y(\omega-\omega_b)\sqrt{\gamma_y}f_y(\omega)]\label{SolBX}	
\end{eqnarray}		
with the cooperativity \cite{Koppenhofer2023}
\begin{equation}
C_o=4g^2/(\kappa\gamma_x)\nonumber
\end{equation}
and susceptibility functions \cite{Clerk2010, Aspelmeyer2014}
\begin{subequations}
\begin{eqnarray}
&&\chi_x(\omega-\omega_b)=\frac{i(\omega-\omega_b)-\frac{\gamma_y}{2}}{[i(\omega-\omega_b)-\frac{\gamma_x}{2}][i(\omega-\omega_b)-\frac{\gamma_y}{2}]+\Omega^2},\nonumber\\
&&\chi_y(\omega-\omega_b)=\frac{1}{i(\omega-\omega_b)-\frac{\gamma_y}{2}}	.\nonumber
\end{eqnarray}
\end{subequations}

\section{Homodyne detection and performance of gyroscop\label{Sec3}}

As a gyroscope, our aim is to readout the unknown angular velocity $\Omega$ through the output $\alpha_{\mathrm{out}}$. Note that the angular velocity $\Omega$ is directly encompassed in the complex amplitude $b_x$, and it is then transferred to the output $\alpha_{\mathrm{out}}$ through coupling with the readout cavity. Therefore, we choose the quadrature $X=x/x_{\mathrm{zpf}}=(b_x+b^\dag_x), \big(x_{\mathrm{zpf}}=\sqrt{\hbar/(2m\omega_b)}\big)$ as the observable quantity to be measured.

As we mentioned, there are three important indices for a gyroscope: 1. \textbf{range}, which determines the range within which the angular velocity $\Omega$ can be readout; 2. \textbf{SNR}, which determines whether the output signal can be read; and 3. \textbf{sensitivity or accuracy}, which determines the minimum detectable change of the angular velocity $\Delta\Omega$, similar to the ticks of a ruler. Therefore, the gyroscope is more sensitive when $\Delta\Omega$ is smaller. In this work, we use $\Delta\Omega^2$ as the sensitivity because the angular velocity always appears squared in all quantities. All of these indices are closely related to the power spectral density of the noise. Thereby, we first introduce the noise power spectral density and then discuss these indices in detail.

\subsection{Noise power spectral density and standard quantum limit}
We now provide a general form of the noise power spectral density for an observable quantity $O$
\begin{eqnarray}
N_{O}(\omega)=\int\mathrm{d}t~e^{i\omega t} [\langle O(t)O(0)\rangle-\langle O(t)\rangle\langle O(0)\rangle]\nonumber\\
=\frac{1}{2\pi}\int\mathrm{d}\omega'~[\langle O(\omega)O(\omega')\rangle-\langle O(\omega)\rangle\langle O(\omega')\rangle]\label{DefineN},
\end{eqnarray}
where the first term on the right-hand side represents the sum of the quantum average and the statistical average, while the second term represents the statistical average alone. The physics behind Eq. \eqref{DefineN} is explained by the Wiener-Khinchin theorem: the autocorrelation function of an observable quantity is connected to its power spectral density through a Fourier transformation \cite{Aspelmeyer2014}. The area under the spectral density equals the fluctuation of the quantity:
\begin{equation}
\frac{1}{2\pi}\int\mathrm{d}\omega~N_{O}(\omega)=\langle O^2(0)\rangle-\langle O(0)\rangle ^2.
\label{WienerKhinchin}
\end{equation}
Note that our definition \eqref{DefineN} differs slightly from those in Ref. \cite{Clerk2010, Aspelmeyer2014}, since our input $\alpha_{\mathrm{in}}$ contains a classical amplitude $\alpha$ besides the quantum  $a_{\mathrm{in}}$. One can see Refs. \cite{Clerk2010, Aspelmeyer2014} for more detail about noise spectral density. In this work, we also consider the symmetric noise power spectral density $\bar{N}_{O}(\omega) =(N_{O}(\omega)+N_{O}(-\omega))/2$.

\subsubsection{$a_{\mathrm{in}}$ in the vacuum state \label{VacuumInput}}
Before we proceed, we now introduce the correlations of the quantum input $a_{\mathrm{in}}(\omega)$
\begin{eqnarray}
&&\langle 0|a_{\mathrm{in}}(\omega)|0\rangle=\langle0|a^\dag_{\mathrm{in}}(\omega)|0 \rangle=0,\nonumber\\
&&\langle 0|a^\dag_{\mathrm{in}}(\omega)a_{\mathrm{in}}(\omega')|0\rangle=0,\nonumber\\
&&\langle 0|a_{\mathrm{in}}(\omega)a^\dag_{\mathrm{in}}(\omega')|0\rangle=2\pi\delta(\omega+\omega').\label{VacuumCorrelation}
\end{eqnarray}

According to the definition \eqref{DefineN}, we can obtain the noise power spectral density
\begin{eqnarray}
N_{X}(\omega)=N_{X}^\mathrm{zpf}(\omega)+N_{X}^\mathrm{ba}(\omega)+N_{X}^\Omega(\omega),
\end{eqnarray}
of the quadrature $X$ using the above correlations, where the first term $N_{X}^\mathrm{zpf}(\omega)=\gamma_x|\chi_x(\omega-\omega_b)|^2$ is the zero-point noise caused by the thermal noise $f_x$ at  zero Kelvin, the second term $N_{X}^\mathrm{ba}(\omega)=\gamma_xC_o|\chi_x(\omega-\omega_b)-\chi_x(\omega+\omega_b)|^2$ represents the back-action noise caused by the quantum fluctuation input $a_{\mathrm{in}}$, and the third term $N_{X}^\Omega(\omega)=\Omega^2\gamma_y|\chi_x(\omega-\omega_b)|^2|\chi_y(\omega-\omega_b)|^2$ denotes the angular noise induced by the thermal noise $f_y$. Furthermore, the total noise power spectral density of quadrature $X$ also includes imprecision noise resulting from the measurement \cite{Clerk2010}, which is associated with the photon current
\begin{eqnarray}
I(\omega)&&=\alpha_{\mathrm{out}}(\omega)+\alpha^\dag_{\mathrm{out}}(\omega)\nonumber\\	
&&=-2\sqrt{\gamma_xC_o}X(\omega)-(\alpha_{\mathrm{in}}(\omega)+\alpha^\dag_{\mathrm{in}}(\omega)).\label{PhotonCurrent}
\end{eqnarray}
We see that the photon current amplifies the  mode $x$ with a coefficient $G=2\sqrt{\gamma_xC_o}$. Therefore, the noise power spectral density of the photon current should also include an amplification for $N_{X}(\omega)$, i.e.,
\begin{eqnarray}
N_{I}(\omega)=&&1+4\gamma_xC_o N_{X}(\omega)\nonumber\\
&&+2\gamma_x C_o[\chi_x(\omega-\omega_b)-\chi_x(\omega+\omega_b)\nonumber\\
&&+\chi_x(-\omega+\omega_b)-\chi_x(-\omega-\omega_b)],\label{VacuumNI}
\end{eqnarray}
where the constant term is the shot noise resulting from the autocorrelation of the quantum input $a_{\mathrm{in}}$ in Eq. \eqref{PhotonCurrent}, the second term represents the amplified $N_{X}(\omega)$, and the third term denotes the noise resulting from the cross-correlation between the quadrature $X$ and the quantum fluctuation input $a_{\mathrm{in}}$. Note that  the third term in Eq. \eqref{VacuumNI} cancels out in the symmetric form,i.e.,
\begin{eqnarray}
\bar{N}_{I}(\omega)=1+4\gamma_x C_o \bar{N}_{X}(\omega)\label{Noise}.
\end{eqnarray}
 Therefore, the total noise power spectral density (symmetric) referred back to quadrature $X$ reads
 \begin{eqnarray}
 \bar{N}_{X}^\mathrm{tot}(\omega)=\bar{N}_{X}^\mathrm{zpf}(\omega)+\bar{N}_{X}^\mathrm{add}(\omega)+\bar{N}_{X}^\Omega(\omega)
 \label{TotalNXX}
 \end{eqnarray}
 with the additional noise
 \begin{equation}
  \bar{N}_{X}^\mathrm{add}(\omega)=\bar{N}_{X}^\mathrm{ba}(\omega)+\bar{N}_{X}^\mathrm{im},\label{VacuumAdd}
 \end{equation}
where $\bar{N}_{X}^\mathrm{im}=1/G^2$ is the imprecision noise mentioned before.

 What we emphasized in Eq. \eqref{TotalNXX} is that the total noise power spectral density is limited by the  standard quantum limit $\bar{N}_{X}^\mathrm{add}(\omega)\geq\bar{N}_{X}^\mathrm{zpf}(\omega)$. In short, the standard quantum limit describes the minimum additional noise allowed by the Heisenberg uncertainty relation \cite{Clerk2010,Aspelmeyer2014}. In addition, although both the angular noise $\bar{N}_{X}^\Omega(\omega)$ and the back-action noise $\bar{N}_{X}^\mathrm{ba}(\omega)$ result from the coupling between the $x$ mode and other modes, the back-action noise $\bar{N}_{X}^\mathrm{ba}(\omega)$ is absorbed into the additional noise $\bar{N}_{X}^\mathrm{add}(\omega)$ but not the angular noise $\bar{N}_{X}^\Omega(\omega)$. The reason is that the angular noise $\bar{N}_{X}^\Omega(\omega)$ is zero when $\Omega=0$ but the back-action noise $\bar{N}_{X}^\mathrm{ba}(\omega)$ is not. Therefore, the angular noise $\bar{N}_{X}^\Omega(\omega)$ does not impose any limitations on the minimum total noise $\bar{N}_{X}^\mathrm{tot}(\omega)$.

Since the minimum total noise spectral density $\bar{N}_{X}^\mathrm{tot}(\omega)$ is obtained at $\omega=\omega_b$, we focus on this frequency and discuss the standard quantum limit in detail. Here, we consider the thermal decay rate $\gamma_{x(y)}$ to be much less than the mechanical frequency $\omega_b$, i.e., a low-damped mechanical oscillator, so that the noise power spectral densities can be approximated as
\begin{subequations}
\begin{eqnarray}
&&\bar{N}_{X}^\mathrm{zpf}(\omega_b)\approx\frac{\gamma_x}{2}\frac{\frac{\gamma_y^2}{4}}{(\frac{\gamma_x\gamma_y}{4}+\Omega^2)^2},\label{Nxxzpf}\\
&&\bar{N}_{X}^\mathrm{add}(\omega_b)\approx\gamma_xC_o\frac{\frac{\gamma_y^2}{4}}{(\frac{\gamma_x\gamma_y}{4}+\Omega^2)^2}+\frac{1}{G^2},\label{Nxxadd}\\
&&\bar{N}_{X}^\Omega(\omega_b)\approx\frac{2\Omega^2}{\gamma_y}\frac{\frac{\gamma_y^2}{4}}{(\frac{\gamma_x\gamma_y}{4}+\Omega^2)^2}.\label{NxxOmega}
\end{eqnarray}
\end{subequations}
One can easily check
\begin{equation}
\bar{N}_{X}^\mathrm{add}(\omega_b)\geq\frac{\frac{\gamma_y}{2}}{(\frac{\gamma_x\gamma_y}{4}+\Omega^2)},\nonumber
\end{equation}
and the equality holds if and only if  $C_o=(\frac{\gamma_x\gamma_y}{4}+\Omega^2)/(\gamma_x\gamma_y)$. In doing so, we  have
\begin{eqnarray}
\bar{N}_{X}^\mathrm{add}(\omega_b)-\bar{N}_{X}^\mathrm{zpf}(\omega_b)&&=\frac{1}{2}\frac{\gamma_y}{\frac{\gamma_x\gamma_y}{4}+\Omega^2}\big(1-\frac{1}{1+\frac{4\Omega^2}{\gamma_x\gamma_y}}\big)\nonumber\\
&&\geq\frac{1}{4}\frac{\gamma_y}{\frac{\gamma_x\gamma_y}{4}+\Omega^2}, ~~(\Omega\neq0)\nonumber
\end{eqnarray}
and thus in this case we find that the system fails to reach the standard quantum limit when the platform is rotating.

Also, one can see that $\bar{N}_{X}^\Omega(\omega_b)\leq\bar{N}_{X}^\mathrm{zpf}(\omega_b)$ when $\Omega^2\leq\gamma_x\gamma_y/4$. This implies that the angular noise $\bar{N}_{X}^\Omega(\omega_b)$  can be disregarded if the angular velocity $\Omega$ is sufficiently small. In other words, the additional noise $\bar{N}_{X}^\mathrm{add}(\omega_b)$ always dominates the total noise $\bar{N}_{X}^\mathrm{tot}(\omega_b)$. This corresponds to a very practical scenario: if the rotation of the platform is much less than that of the Earth (about $10^{-5}~\mathrm{rad/s}\approx10^{-6}~\mathrm{Hz}$), the total noise $\bar{N}_{X}^\mathrm{tot}(\omega)$ is almost independent of the angular velocity $\Omega$ for mechanical oscillators with quality factors $Q\approx~10^{3}-10^{5}$ and operating frequencies $\mathrm{kHz}-\mathrm{MHz}$ \cite{Aspelmeyer2014}. Therefore, it would be impossible to readout the angular velocity from the noise spectral density, as was done in Ref. \cite{Li2018B1}. In addition, in Ref. \cite{Li2018B1} the authors also ignore the constraints of the standard quantum limit on the total noise power spectral density $\bar{N}_{X}^\mathrm{tot}(\omega)$. Therefore, the assumption they considered $\bar{N}_{X}^\mathrm{add}(\omega_b)\ll\bar{N}_{X}^\mathrm{zpf}(\omega_b)$ would not hold true in experiments when the quantum input is in a vacuum state.

\subsubsection{$a_{\mathrm{in}}$ in the squeezed vacuum state}
In this case, we consdier $a_{\mathrm{in}}$ in a single-mode squeezed vacuum state
\begin{equation}
|\xi\rangle=S(\xi)|0\rangle=e^{-\frac{1}{2}(\xi a^\dag_{\mathrm{in}}(\omega)a^\dag_{\mathrm{in}}(\omega)-\xi^*a_{\mathrm{in}}(\omega)a_{\mathrm{in}}(\omega))}|0\rangle
\end{equation}
with an arbitrary complex number $\xi=r e^{i\phi}, (r>0)$. Correspondingly, the correlations of the quantum input $a_{\mathrm{in}}$ are calculated by
\begin{eqnarray}
&&\langle \xi|a_{\mathrm{in}}(\omega)|\xi\rangle=\langle\xi|a^\dag_{\mathrm{in}}(\omega)|\xi\rangle=0,\nonumber\\
&&\langle\xi|a_{\mathrm{in}}(\omega)a^\dag_{\mathrm{in}}(\omega')|\xi\rangle=2\pi\delta(\omega+\omega')\cosh^2 r,\nonumber\\
&&\langle\xi|a^\dag_{\mathrm{in}}(\omega)a_{\mathrm{in}}(\omega')|\xi\rangle=2\pi\delta(\omega+\omega')\sinh^2 r.\label{SqueezeCorrelation}
\end{eqnarray}
In doing so, the symmetric noise spectral density of the photon current reads
\begin{equation}
\bar{N}_{I}(\omega)=e^{-2r}+4\gamma_x C_o \bar{N}_{X}(\omega)\label{SqueezedNI}
\end{equation}
with
\begin{equation}
\bar{N}_{X}(\omega)=\bar{N}_{X}^\mathrm{zpf}(\omega)+e^{-2r}\bar{N}_{X}^\mathrm{ba}(\omega)+\bar{N}_{X}^\Omega(\omega),
\end{equation}
where we have assumed $\phi=\pi$ in the derivation to obtain an attenuated shot noise, i.e., the first term in Eq. \eqref{SqueezedNI}. Accordingly, the total noise spectral density for the mechanical $x$ mode becomes
\begin{equation}
\bar{N}^{\mathrm{tot}}_{X}(\omega)=\bar{N}_{X}^\mathrm{zpf}(\omega)+\bar{N}_{X}^{\mathrm{add},s}(\omega)+\bar{N}_{X}^\Omega(\omega)
\end{equation}
with the squeezed additional noise
\begin{equation}
\bar{N}_{X}^{\mathrm{add},s}(\omega)=e^{-2r}\bar{N}_{X}^\mathrm{ba}(\omega)+\bar{N}_{X}^\mathrm{im}.\label{SqueezeAdd}	
\end{equation}

Compared to Eq. \eqref{VacuumAdd}, we see that the squeezed vacuum input causes the back-action noise $\bar{N}_{X}^\mathrm{ba}(\omega)$ decrease exponentially, but it does not affect the imprecision noise $\bar{N}_{X}^\mathrm{im}$. This provides an opportunity to achieve or surpass the standard quantum limit. For clear comparison, we also specialize to the frequency $\omega=\omega_b$ as did in Sec. \ref{VacuumInput} and then  have
\begin{equation}
\bar{N}_{X}^{\mathrm{add},s}(\omega_b)\geq e^{-r}\frac{\frac{\gamma_y}{2}}{(\frac{\gamma_x\gamma_y}{4}+\Omega^2)},\nonumber
\end{equation}
where the equality holds if and only if $C_o=e^r (\frac{\gamma_x\gamma_y}{4}+\Omega^2)/(\gamma_x\gamma_y)$. At this time, we further obtain
\begin{eqnarray}
\bar{N}_{X}^{\mathrm{add},s}(\omega_b)-\bar{N}_{X}^\mathrm{zpf}(\omega_b)\geq\frac{1}{2}\frac{\gamma_y}{\frac{\gamma_x\gamma_y}{4}+\Omega^2}(e^{-r}-\frac{1}{2}), (\Omega\neq0),\nonumber
\end{eqnarray}
 so that the system can reach and even surpass the standard quantum limit \cite{Clerk2010,Aspelmeyer2014}  when the condition $r\geq\ln{2}$ is satisfied.

Re-examining the noise spectral density from the perspective of the output, i.e., the noise spectral density $\bar{N}_{I}(\omega_b)$, we can easily find that the zero-point noise $\bar{N}^{\mathrm{zpf}}_{X}(\omega_b)$ dominates the noise spectral density $\bar{N}_{I} (\omega_b)$ when the conditions $r\rightarrow\infty$ and $\Omega^2\leq\gamma_x\gamma_y/4$ are satisfied. If we consider the shot noise as the floor of the noise spectral density of the photon current under the vacuum input (cf. Eq. \eqref{Noise}), this floor aproaches zero with squeezing. As a result, the noise spectral density $\bar{N}_{I} (\omega_b)$ only depends on the the zero-point noise $\bar{N}^{\mathrm{zpf}}_{X}(\omega_b)$. In other words, the noise resulting from the quantum input $a_{\mathrm{in}}$ can be eliminated by squeezing, so that the system is only influenced by the zero-point noise $\bar{N}^{\mathrm{zpf}}_{X}(\omega_b)$. This is the major difference of noise power spectral densities between the vacuum and squeezed vacuum inputs.

\subsection{Signal-to-noise ratio, range and sensitivity}
After introducing the noise spectral density in the previous section, we now analyze range, SNR, and sensitivity of the gyroscope. Using the correlations \eqref{VacuumCorrelation} and \eqref{SqueezeCorrelation}, we define the signal specrtrum as
\begin{eqnarray}
S(\omega)=&&|\langle I(\omega)\rangle-\langle I(-\omega)\rangle|^2\nonumber\\
=&&|2\gamma_xC_o[\chi_x(\omega-\omega_b)+\chi_x^*(\omega-\omega_b)\nonumber\\
&&-\chi_x(\omega+\omega_b)-\chi_x^*(\omega+\omega_b)](\alpha-\alpha^*)|^2\nonumber\\
=&&16N_{\mathrm{in}}\gamma_x^2C_o^2|\chi_x(\omega-\omega_b)+\chi_x^*(\omega-\omega_b)\nonumber\\
&&-\chi_x(\omega+\omega_b)-\chi_x^*(\omega+\omega_b)|^2,
\end{eqnarray}
where we set $\varphi=\mathrm{arg}~\alpha=\pi/2$ for brevity. As the statistical average of the photon current, the output signal only depends on the susceptibility function of the system and is proportional to the input photon number. At the frequency $\omega=\omega_b$, it can be approximated as
\begin{equation}
S(\omega_b)\approx 16N_{\mathrm{in}}C_o^2\frac{\gamma_x^2\gamma_y^2}{(\frac{\gamma_x\gamma_y}{4}+\Omega^2)^2},\label{SIG}
\end{equation}
where we also assume $(\Omega,\gamma_{x(y)})\ll\omega_b$ as we did when deriving Eqs. \eqref{Nxxzpf}-\eqref{NxxOmega}. Once the gyroscope is designed, one can use this equation to readout the angular velocity $\Omega$ from the measured signal.

With the signal spectrum $S(\omega)$ and the noise spectral density $\bar{N}_{I}(\omega)$, one can easily write the SNR as
\begin{equation}
\mathrm{SNR}(\omega)=\frac{S(\omega)}{\bar{N}_{I}(\omega)}.
\end{equation}
When the quantum input $a_{\mathrm{in}}$ is in the vacuum state or the squeezed vacuum state, SNR per photon can be approximated as
\begin{equation}
\frac{\mathrm{SNR}_{v}(\omega_b)}{N_{\mathrm{in}}}\approx\frac{16C_o^2\frac{\gamma_x^2\gamma_y^2}{(\frac{\gamma_x\gamma_y}{4}+\Omega^2)^2}}{1+C_o(C_o+2\frac{\frac{\gamma_x\gamma_y}{4}+\Omega^2}{\gamma_x\gamma_y})\frac{\gamma_x^2\gamma_y^2}{(\frac{\gamma_x\gamma_y}{4}+\Omega^2)^2}}
\end{equation}
or
\begin{eqnarray}
\frac{\mathrm{SNR}_{s}(\omega_b)}{N_{\mathrm{in}}}\approx\frac{16e^{2r}C_o^2\frac{\gamma_x^2\gamma_y^2}{(\frac{\gamma_x\gamma_y}{4}+\Omega^2)^2}}{1+C_o(C_o+2\frac{e^{2r}\frac{\gamma_x\gamma_y}{4}+\Omega^2}{\gamma_x\gamma_y})\frac{\gamma_x^2\gamma_y^2}{(\frac{\gamma_x\gamma_y}{4}+\Omega^2)^2}},\nonumber\\
\end{eqnarray}
where the subscripts $v$ and $s$ label the vacuum and the squeezed vacuum, respectively. It should be pointed out that the output signal can only be readout when the SNR per photon is greater than 1. The advantage of this is that the readability of the output signal does not depend on the pump power of the input field but only relates to the system parameters. In doing so, it provides an upper bound for the angular velocity and the lower bound for the cooperativity. When the quantum input $a_{\mathrm{in}}$ is in the vacuum state, these bounds read
\begin{equation}
0\leq\Omega_v^2\leq(3C_{o}-\frac{1}{4})\gamma_x\gamma_y
\label{RangeOmega}
\end{equation}
and
\begin{equation}
C_{o}\geq\frac{1}{12}.
\label{RangeCo}	
\end{equation}
When the quantum input $a_{\mathrm{in}}$ is in the squeezed vacuum state, they become
\begin{eqnarray}
0\leq\Omega_s^2\leq(\sqrt{e^{4r}+16e^{2r}-1}C_o-e^{2r}C_o-\frac{1}{4})\gamma_x\gamma_y
\label{SqueezeRangeOmega}
\end{eqnarray}
and
\begin{equation}
C_o\geq\frac{1}{4}\frac{1}{\sqrt{e^{4r}+16e^{2r}-1}-e^{2r}}.
\label{SqueezedRangeCo}	
\end{equation}
One can easily check that the conditions \eqref{SqueezeRangeOmega} and \eqref{SqueezedRangeCo} reduce to those \eqref{RangeOmega} and \eqref{RangeCo} when the squeezed parameter $r=0$.

After finishing the discussions on the range and the SNR, we now focus on the sensitivity \cite{Wu2020,Guo2021,Koppenhofer2023} which is given as
\begin{equation}
\Delta\Omega^2(\omega)=\sqrt{\frac{\bar{N}_{I}(\omega)}{|\partial_{\Omega^2}\big(\langle I(\omega)\rangle-\langle I(-\omega)\rangle\big)|^2}}.
\label{Sensitivity}
\end{equation}
Once again, we emphasize that sensitivity refers to the minimum detectable variation of the angular velocity. Therefore, the gyroscope is more sensitive when $\Delta\Omega^2$ is smaller. Mathematically, this definition states that the product $(\Delta\Omega^2)^2\times|\partial_{\Omega^2}\big(\langle I(\omega)\rangle-\langle I(-\omega)\rangle\big)|^2$ is equal to the noise of photon current $\bar{N}_{I}(\omega)$. Note that the noise also represents the fluctuation (cf. Eq. \eqref{WienerKhinchin}), so that Eq.\eqref{Sensitivity}   physically means that the fluctuation $\bar{N}_{I}(\omega)$ as the minimum change of the output photon current, functions as the reference for observing variations of angular velocity.

When the quantum input $a_{\mathrm{in}}$ is in the vacuum state, Eq. \eqref{Sensitivity} can be approximated as follows at the frequency $\omega=\omega_b$
\begin{equation}
\Delta \Omega_v^2(\omega_b)\approx\frac{\frac{\gamma_x\gamma_y}{4}+\Omega^2}{4\sqrt{N_{\mathrm{in}}}}(1+\frac{\frac{\gamma_x\gamma_y}{4}+\Omega^2}{C_o\gamma_x\gamma_y}).
\label{SensitivityVacuum}
\end{equation}
Note that the sensitivity is proportional to  the thermal decay rates, and thus it tends to zero as the thermal decay rates tend to zero. In other words, the system can distinguish the infinitesimal changes of angular velocity in this extreme case. Physically, this extreme case means that the output photon current is extremely sensitive to the change of the angular velocity when the system is only affected by the shot noise.

It is not hard to examine that the sensitivity has a fundamental limit
\begin{equation}
\Delta\Omega_v^2(\omega_b)\geq\frac{\frac{\gamma_x\gamma_y}{4}+\Omega^2}{2\sqrt{N_{\mathrm{in}}}},
\label{VacuumSensitivityLimit}
\end{equation}
where the equality holds if and only if $C_o=(\frac{\gamma_x\gamma_y}{4}+\Omega^2)/(\gamma_x\gamma_y)$. Notably, the condition $C_o\rightarrow\infty$ indicates that the coupling coefficient $g$ tends to infinity, which is impossible in realistic scenarios. This limitation reduces to the lowest bound of the sensitivity permitted by the standard quantum limit when $\Omega=0$.

When the quantum input $a_{\mathrm{in}}$ is in the squeezed vacuum state, we have
\begin{eqnarray}
&&\Delta \Omega_s^2(\omega_b)\approx e^{-r}\frac{\frac{\gamma_x\gamma_y}{4}+\Omega^2}{4\sqrt{N_{\mathrm{in}}}C_o\gamma_x\gamma_y}\nonumber\\
&&\times\sqrt{(\frac{\gamma_x\gamma_y}{4}+\Omega^2+e^{2r}C_o\gamma_x\gamma_y)^2+(1-e^{4r})C_o^2\gamma_x^2\gamma_y^2}.\nonumber\\
\label{SqueezedSensitivity}
\end{eqnarray}
Also, it gives the limitation
\begin{equation}
\Delta\Omega_s^2(\omega_b)\geq\sqrt{2(1+e^{-2r})}\frac{(\frac{\gamma_x\gamma_y}{4}+\Omega^2)^{3/2}}{4\sqrt{N_{\mathrm{in}}C_o\gamma_x\gamma_y}},
\label{SqueezedSensitivityLimit}
\end{equation}
where the equality holds if and only if $C_o=(\frac{\gamma_x\gamma_y}{4}+\Omega^2)/(\gamma_x\gamma_y)$. The most significant point is that the sensitivity can be improved by squeezing but the improvement is limited:
\begin{eqnarray}
\frac{\Delta\Omega_s^2(\omega_b)}{\Delta\Omega_v^2(\omega_b)}=&&\sqrt{e^{-2r}+\frac{2(1-e^{-2r})C_o\gamma_x\gamma_y(\frac{\gamma_x\gamma_y}{4}+\Omega^2)}{(\frac{\gamma_x\gamma_y}{4}+\Omega^2+C_o\gamma_x\gamma_y)^2}}\nonumber\\
&&\leq\frac{\sqrt{2}}{2}\sqrt{1+e^{-2r}}.\label{SensitivityComparison}
\end{eqnarray}
The inequality in the second line shows that squeezing is not a very effective method for improving sensitivity, as sensitivity can only be enhanced up to $\sqrt{2}/2$ even when the squeezed parameter $r$ approaches infinity.

\begin{figure} [bt]
\centering
\includegraphics[width=0.48\textwidth]{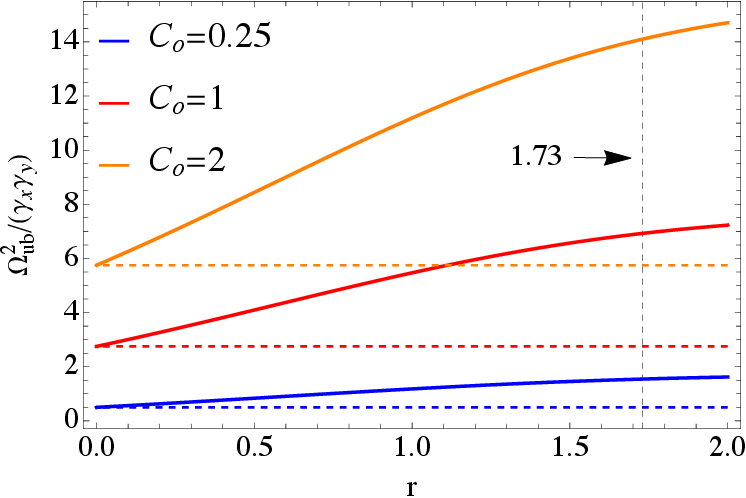}
\caption{(Color online) Numerical simulations of the range $\Omega^2$ under different cooperativities $C_o$. The solid (dashed) lines represent the quantum input $a_{\mathrm{in}}$ in the squeezed vacuum (vacuum) state, and the grid-line is used to mark the accessible highest squeezed parameter $r=1.73$ in experiments \cite{Vahlbruch2016}. The differences between solid and dashed lines demonstrate that squeezing is an effective way for expanding the upper bounds of the angular velociy $\Omega^2$. Also, the degree of this extent of this expansion increases as the squeezed parameter $r$ increases. }
\label{RangeOmegaFigure}
\end{figure}

\section{Numrical Results and Further Discussion\label{Sec4}}

\begin{figure*} [ht]
\centering
\includegraphics[width=0.96\textwidth]{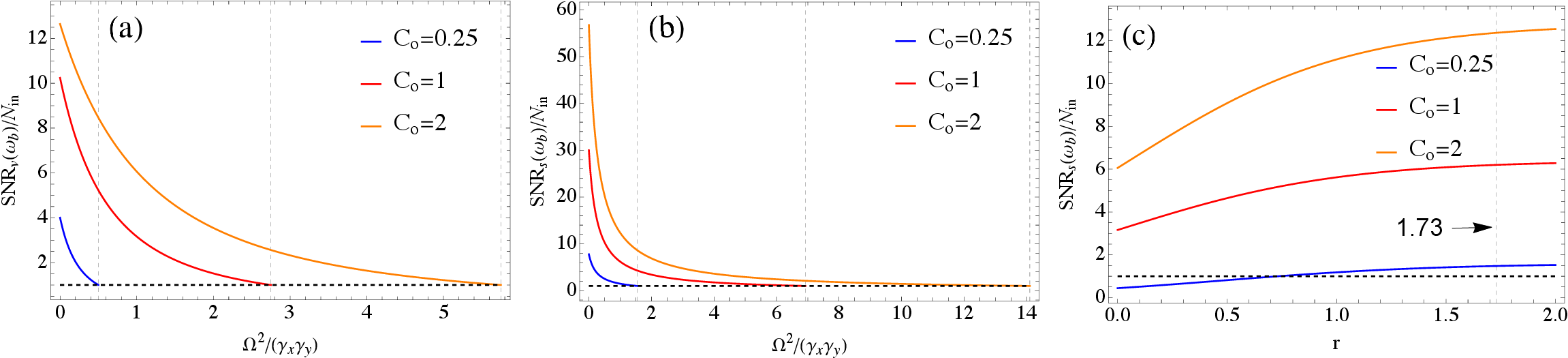}
\caption{(Color online) Numerical simulations of the signal-to-noise ratio per photon $\mathrm{SNR}(\omega_b)/N_{\mathrm{in}}$ at various cooperativities $C_o$. The black dashed line represents the critical condition for  readability $\mathrm{SNR}(\omega_b)=1$. The gridlines are used to label the upper bounds of the angular velocity $\Omega_{\mathrm{ub}}^2$ in (a) and (b), and to label the accessible highest squeezed parameter $r$ in (c). (a) The quantum input $a_{\mathrm{in}}$ is in the vacuum state. (b) The quantum input $a_{\mathrm{in}}$ is in the squeezed vacuum state, where we have fixed  the squeezed parameter $r=1.73$. Compare (a) with (b), one can see that the signal-to-noise ratio per photon is significantly improved under the same cooperativity $C_o$ after squeezing. (c) $\mathrm{SNR}_s(\omega_b)/N_{\mathrm{in}}$ as functions of the squeezed parameter $r$ with a fixed $\Omega^2=\gamma_x\gamma_y$. We observe that the blue solid line is not consistently higher than the black dashed line. Therefore, the squeezed parameter $r$ cannot be too small to ensure the readability when the cooperativity $C_o=0.25$.}
\label{SNRFigure}
\end{figure*}

So far, we have obtained constraints of the cooperativity and upper bounds of the angular velocity at the frequency $\omega=\omega_b$. Based on these constraints, we now simulate the SNR and sensitivity for further discussion. To this end, we first need to determine the range of the angular velocity according to Eqs. \eqref{RangeOmega} and \eqref{SqueezeRangeOmega}. However, the range receives little attention as one of the most important parameters of gyroscopes in recent works \cite{Davuluri2016, Davuluri2017, Li2018, Li2018B1, Li2018C1, Lavrik2019, Wu2020,Guo2021}.

Before discussing the range, we also need to limit the squeezed parameter $r$. At present, the highest accessible degree of squeezing is $15~\mathrm{dB}$ ($15=10~\log_{10}(e^{2r})$, corresponds to $r\approx1.73$) in experiments \cite{Vahlbruch2016}. So that the condition $C_o\geq\frac{1}{12}$ is also valid for Eq. \eqref{SqueezedRangeCo} when $r\in[0,1.73]$.

The numerical simulation of the upper bound of the angular velocity $\Omega^2$ as a function of the squeezed parameter $r$ is shown in Fig. \ref{RangeOmegaFigure}. Here, the subscript ub is an abbreviation for the upper bound of the inequalities \eqref{RangeOmega} and \eqref{SqueezeRangeOmega}. The solid and dashed lines represent the cases of quantum fluctuation input $a_\mathrm{in}$ in the squeezed vacuum state and the vacuum state, respectively. Moreover, the cooperativity $C_o$ is limited in the vicinity of the impedance matching condition $C_o=1$, which is widely utilized in experiments \cite{Koppenhofer2023}. Also, we use the gridline with $r=1.73$ to mark the highest achievable squeezed degree in experiments. The differences between the solid and dashed lines demonstrate that squeezing can enhance the upper bound of the angular velocity $\Omega^2$ by up to 2 times within the accessible range of the squeezed parameter $r$. Moreover, the upper bound of the angular velocity $\Omega^2$ increases monotonically with increasing cooperativity $C_o$, whether the quantum input $a_{\mathrm{in}}$ is in the vacuum state or the squeezed vacuum state. Note that the cooperativity $C_o$ represents the coupling strength between the quantity to be measured and the readout cavity, and thus maintaining a higher coupling strength in experiments can result in a larger range for the gyroscope.

Based on the upper bounds of the angular velocity $\Omega^2$ mentioned above, we simulate SNR under different cooperativities $C_o$, as depicted in Fig. \ref{SNRFigure}. Here, we plot the case of the quantum fluctuation input $a_{\mathrm{in}}$ in the vacuum state in Fig. \ref{SNRFigure} (a), and plot the case of $a_{\mathrm{in}}$ in the squeezed vacuum state in Figs. \ref{SNRFigure} (b) and (c). Also, we use gridlines to mark the upper bounds of the angular velocity, and we use black dashed lines to label the critical readable condition $\mathrm{SNR}(\omega_b)=1$. From the Figs. \ref{SNRFigure} (a) and (b), one can see that the SNR per photon $\mathrm{SNR}(\omega_b)/N_{\mathrm{in}}$ monotonously increases with incresasing cooperativity $C_o$, where we fixed the squeezed parameter $r=1.73$ in Fig. \ref{SNRFigure} (b). Therefore,  maintaining a higher cooperativity $C_o$ can not only obtain a wider range but also can improve readability. This result extends the findings presented in Fig. \ref{RangeOmegaFigure}. One can also see that the SNR per photon monotonically decreases as the angular velocity increases. This result shows that readability improves when the angular velocity is reduced, especially when the quantum fluctuation input $a_{\mathrm{in}}$ is squeezed. In Fig. \ref{SNRFigure} (c), we replot the SNR per photon $\mathrm{SNR}_s(\omega_b)/N_{\mathrm{in}}$ as a function of the squeezed parameter $r$ under different cooperativities $C_o$, where we fixed the angular velocity $\Omega^2=\gamma_x\gamma_y$. Under different cooperativities, all three curves of SNR per photon  monotonically increase with increasing squeezing parameter. Therefore, a higher degree of squeezing is beneficial for further improving the readability of the gyroscope.

\begin{figure*} [bt]
\centering
\includegraphics[width=0.96\textwidth]{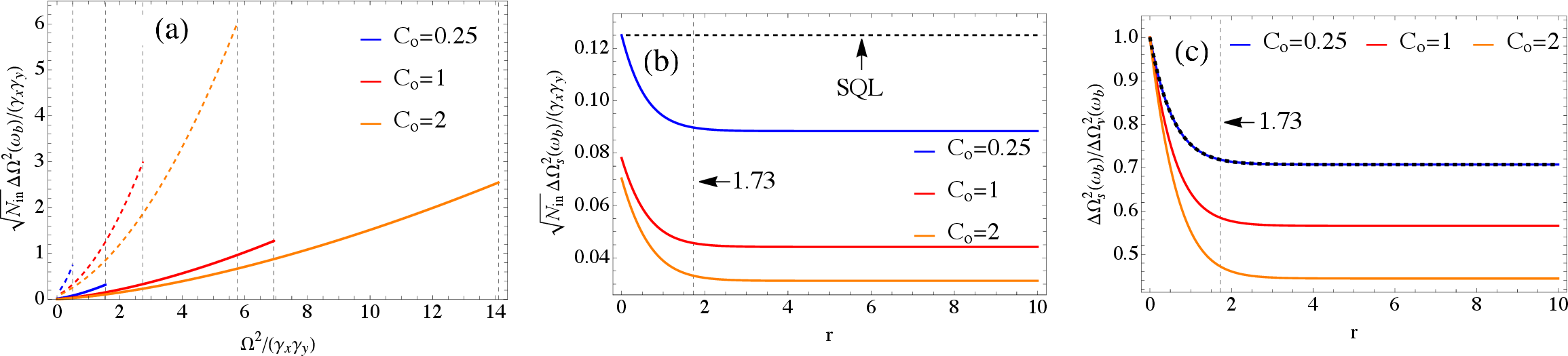}
\caption{(Color online) The numerical simulation of the sensitivity $\sqrt{N_{\mathrm{in}}}\Delta\Omega^2(\omega_b)$. In (a), the solid (dashed) lines represent the case of the quantum fluctuation input $a_{\mathrm{in}}$ in the squeezed vacuum (vacuum) state. Gridlines are used to label the corresponding upper bounds of the angular velocity $\Omega_{\mathrm{ub}}^2$. Also, we set the squeeze parameter $r=1.73$ in plotting solid lines. Under the same cooperativity $C_o$, one can observe that $\sqrt{N_{\mathrm{in}}}\Delta\Omega^2(\omega_b)$ decreases significantly after squeezing. In (b), we plot $\sqrt{N_{\mathrm{in}}}\Delta\Omega^2(\omega_b)$ as functions of the squeezed parameter $r$ with the fixed angular velocity $\Omega=0$ in solid lines. The black dashed line marks the standard quantum limit obtained from Eq. \eqref{VacuumSensitivityLimit}. One can see that all three solid lines are lower than the black dashed line, indicating that squeezing is an effective method for surpassing the standard quantum limit. In addition, $\sqrt{N_{\mathrm{in}}}\Delta\Omega_s^2(\omega_b)$ decreases as the squeezed parameter $r$ increases and eventually approaches its limitation \eqref{SqueezedSensitivityLimit}. This indicates that squeezing is not an effective method for further improving the sensitivity. In (c), for the sake of clear comparison with the case of the quantum fluctuation input $a_{\mathrm{in}}$ in the vacuum state, we replot the ratio $\Delta\Omega_s^2(\omega_b)/\Delta\Omega_v^2(\omega_b)$ as functions of the squeezed parameter $r$ with the fixed angular velocity $\Omega=0$. Here, the black dashed line is plotted based on the inequality of Eq. \eqref{SensitivityComparison}. We can clearly see that the sensitivity can only be improved by up to $\sqrt{2}/2$ after squeezing.}
\label{SensitivityFigure}
\end{figure*}

After finishing the discussions on the range and SNR per photon, we now shift our focus to sensitivity, which is the most crucial index for gyroscopes. The corresponding numerical simulation is shown in Fig. \ref{SensitivityFigure}. In Fig. \ref{SensitivityFigure} (a), we use  solid (dashed) lines to represent the case of the quantum fluctuation input $a_{\mathrm{in}}$ in the squeezed vacuum (vacuum) state. We also use gridlines to mark the corresponding upper bounds of the angular velocity $\Omega_{\mathrm{ub}}^2$ for a given cooperativity $C_o$. One can observe that $\sqrt{N_{\mathrm{in}}}\Delta\Omega^2(\omega_b)$ decreases as the cooperativity $C_o$ increases. This implies that the system is capable of discerning smaller changes in angular velocity $\Omega^2$ when the cooperativity $C_o$ is higher. Furthermore, one can also see that $\sqrt{N_{\mathrm{in}}}\Delta\Omega^2(\omega_b)$ monotonously increases with the increasing angular velocity $\Omega^2$. This indicates that our gyroscope is more sensitive to smaller angular velocities. In Fig. \ref{SensitivityFigure} (b), we replot the sensitivity $\sqrt{N_{\mathrm{in}}}\Delta\Omega^2_s(\omega_b)$ as a function of the squeezed parameter $r$ with the fixed angular velocity $\Omega^2=0$. Here, we also use gridlines to mark the highest accessible squeezed parameter $r=1.73$, and we use a black dashed line to label the standard quantum limit obtained from Eq. \eqref{VacuumSensitivityLimit}. In addition, we appropriately extend the squeezed parameter $r$ to observe the limitation \eqref{SqueezedSensitivityLimit}. We observe that the sensitivity $\sqrt{N_{\mathrm{in}}}\Delta\Omega_s^2(\omega_b)$ is consistently lower than the standard quantum limit when the quantum fluctuation input $a_{\mathrm{in}}$ is squeezed. Moreover, $\sqrt{N_{\mathrm{in}}}\Delta\Omega_s^2(\omega_b)$ decreases monotonically with the increasing squeezed parameter $r$ and eventually approaches its limit. Again, this result shows that using a squeezed vacuum state reduces $\sqrt{N_{\mathrm{in}}}\Delta\Omega_s^2(\omega_b)$, but it is not the most effective method. Also, $\sqrt{N_{\mathrm{in}}}\Delta\Omega_s^2(\omega_b)$ decreases as the cooperativity $C_o$ increases. Therefore, maintaining higher cooperativity $C_o$ in experiments is beneficial for improving sensitivity. For the sake of clear comparisons, we plot the ratio $\Delta\Omega_s^2(\omega_b)/\Delta\Omega_v^2(\omega_b)$ as a function of the squeezed parameter $r$ in Fig. \ref{SensitivityFigure} (c). In this plot, we set the angular velocity $\Omega=0$. Here, the solid lines are plotted based on equality while the black dashed line is plotted based on the inequality in Eq. \eqref{SensitivityComparison}. Obviously, the ratio $\Delta\Omega_s^2(\omega_b)/\Delta\Omega_v^2(\omega_b)$ is always lower than $\sqrt{2}/2$. Therefore, $\sqrt{N_{\mathrm{in}}}\Delta\Omega_s^2(\omega_b)$ can only be reduced by a maximum of $\sqrt{2}/2$.

At the end of this section, we provide a brief review of our numerical simulations. Firstly, maintaining a higher cooperativity $C_o$ in experiments is beneficial for obtaining a wider range, a higher SNR per photon, and a higher sensitivity. Secondly, using a squeezed vacuum state as the input reduces the minimum detectable change of the angular velocity, but it decreases by a maximum of $\sqrt{2}/2$.

\section{Conclusion\label{Sec5}}
In conclusion, we have proposed a quantum gyroscope that utilizes a coupled cavity system. We start by discussing the noise power spectral density and delve into the details of the standard quantum limit. We systematically analyze all three crucial indices for gyroscopes: range, signal-to-noise ratio, and sensitivity, instead of focusing on just one of them, as done in recent proposals \cite{Davuluri2016, Davuluri2017, Li2018, Li2018B1, Li2018C1, Wu2020,Guo2021}. Based on this comprehensive analysis, we provide fundamental sensitivity limits for quantum inputs in the vacuum and squeezed vacuum states, respectively. However, the two fundamental limits of sensitivity are the most important index of gyroscopes, yet they have received little attention in the aforementioned proposals, particularly when the quantum input is squeezed. More importantly, we find that squeezing can enhance sensitivity and surpass the standard quantum limit. However, this enhancement can only reach up to $\sqrt{2}/2$ even as the squeezing parameter approaches infinity. This result provides a basis for guiding experiments, indicating that squeezing is not a very effective method for further improving sensitivity.

\section*{Acknowledgements}
 This work is supported by the National Natural Science Foundation of China (NSFC) under Grants No. 62273226, No. 61873162.

\appendix*
\section{Hamiltonian of the double-mode oscillator in a rotating platform\label{Appendix}}

\begin{figure}[t]
\centering
\includegraphics[width=0.48\textwidth]{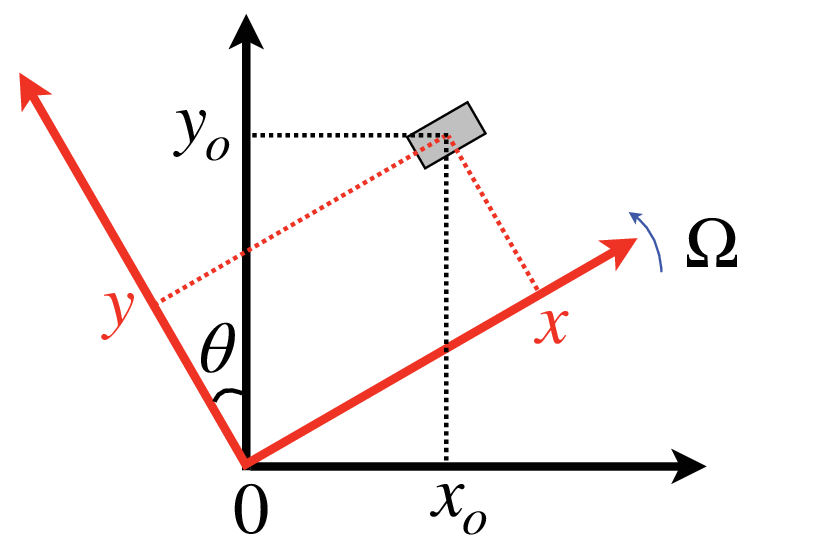}
\caption{(Color online) Schematic of coordinate systems before and after rotation, where the shaded box represents the double-mode oscillator. The $x_o-0-y_o$ coordinate system (inertial system, plotted in black) rotates counterclockwise with an angular velocity $\Omega$. After time $t$, it forms the $x-0-y$ coordinate system (non-inertial system, plotted in red) such that the rotated angle is $\theta=\int_0^t\mathrm{d}\tau~\Omega$.}
\label{Rotation}
\end{figure}

In the appendix, In the appendix, we start by introducing the Lagrangian and then proceed to derive the Hamiltonian of a double-mode mechanical oscillator in a rotating coordinate system. According to the model in Fig. \ref{model}, the coordinate systems before and after rotation are shown in Fig. \ref{Rotation}. The $x_o-y_o$ coordinate system (inertial system, plotted in black) rotates counterclockwise with an angular velocity $\Omega$ and then transforms into the $x-y$ coordinate system (non-inertial system, plotted in red). After a time $t$, the rotated angle is $\theta=\int_0^t\mathrm{d}\tau~\Omega$. In doing so, the positions of the mechanical oscillator (shadow box) in two systems are $(x_o,y_o)$ and $(x,y)$, respectively. The transformation relation between coordinates in these two systems reads
\begin{eqnarray}
\begin{pmatrix}
x\\
y		
\end{pmatrix}
=\begin{pmatrix}
\cos\theta(t) & \sin\theta(t)\\
-\sin\theta(t) & \cos \theta(t)	
\end{pmatrix}
\begin{pmatrix}
x_o\\
y_o
\end{pmatrix}.
\label{TransformationRelation}
\end{eqnarray}
Correspondingly, the Lagrangian of the oscillator in the original system reads
\begin{equation}
\mathcal{L}_o=T_o-V_o\label{LagrangianO}
\end{equation}
with the kinetic energy $T_o$
\begin{equation}
T_o(\dot{x}_o,\dot{y}_o)=\frac{m}{2} (\dot{x}^2_o+\dot{y}^2_o)\nonumber
\end{equation}
and the potential energy $V_o$
\begin{eqnarray}
&&V_o(x_o,y_o)=\frac{1}{2}[k_x(x-x_e)^2+k_y(y-y_e)^2]\nonumber\\
&&=\frac{k_x}{2}[(x_o-x_{oe})\cos\theta(t)+(y_o-y_{oe})\sin\theta(t)]^2\nonumber\\
&&+\frac{k_y}{2}[-(x_o-x_{oe})\sin\theta(t)+(y_o-y_{oe})\cos\theta(t)]^2,\nonumber
\end{eqnarray}
where $k_{x(y)}$ is the spring constant of the $x(y)$ mode. Here, $x_e, y_e$ and $x_{oe}, y_{oe}$ are the equilibrium positions in the rotation system, while $x_{oe}$ and $y_{oe}$ represent the equilibrium positions in the original system. The Lagrangian \eqref{LagrangianO} further gives the Hamiltonian
\begin{eqnarray}
&&H_o(x_o,y_o,p_{o,x},p_{o,y})=\frac{1}{2m}(p_{o,x}^2+p_{o,y}^2)\nonumber\\
&&+\frac{k_x}{2}[(x_o-x_{oe})\cos\theta(t)+(y_o-y_{oe})\sin\theta(t)]^2\nonumber\\
&&+\frac{k_y}{2}[-(x_o-x_{oe})\sin\theta(t)+(y_o-y_{oe})\cos\theta(t)]^2,
\end{eqnarray}
where the momenta $p_{x,o}, p_{y,o}$ conjugate to coordinates $x_o,y_o$ read
\begin{eqnarray}
&&p_{x,o}=\frac{\partial \mathcal{L}_o}{\partial \dot{x}_o}=m\dot{x}_o,\label{PXO}\\
&&p_{y,o}=\frac{\partial \mathcal{L}}{\partial \dot{y}_o}=m\dot{y}_o.\label{PYO}
\end{eqnarray}
In addition, these two sets of conjugate operators satisfy the basic commutation relation $[x_o,p_{x,o}]=[y_o,p_{y,o}]=i\hbar$.

In the rotating system, the Lagrangian of the mechanical oscillator reads
\begin{eqnarray}
\mathcal{L}=&&T(\dot{x},\dot{y})-V(x,y)\nonumber\\
=&&\frac{m}{2}(\dot{x}^2+\dot{y}^2)-\frac{1}{2}[k_x(x-x_o)^2+k_y(y-y_o)^2]\nonumber\\
&&+m\Omega(-\dot{x}y+x\dot{y})+\frac{m\Omega^2}{2}(x^2+y^2),
\label{LagrangianR}
\end{eqnarray}
where the first term in the second line represents the translational kinetic energy, the second term is the potential energy, the third term denotes the energy induced by Coriolis forces, and the last term is the energy induced by the centrifugal forces. The last two terms are fictitious energy that arises from non-inertial forces.

Similarly, this Lagrangian \eqref{LagrangianR} gives the Hamiltonian
\begin{eqnarray}
H(x,y,p_x,p_y)=&&\frac{1}{2m}[(p_x+m\Omega y)^2+(p_y-m\Omega x)^2]\nonumber\\
&&+\frac{1}{2}[k_x(x-x_o)^2+k_y(y-y_o)^2]\nonumber\\
&&-\frac{m\Omega^2}{2}(x^2+y^2),
\label{HamiltonianR}
\end{eqnarray}
where the momenta $p_x,p_y$ conjugate to the coordinates $x,y$ are given by
\begin{subequations}
\begin{eqnarray}
&&p_x=\frac{\partial \mathcal{L}}{\partial \dot{x}}=m\dot{x}-m\Omega y,\label{PXR}\\
&&p_y=\frac{\partial \mathcal{L}}{\partial \dot{y}}=m\dot{y}+m\Omega x\label{PYR}.
\end{eqnarray}
\end{subequations}
The first term in Eq. \eqref{HamiltonianR} is the kinetic energy in the non-inertial system, the second term denotes the potential energy, and the third term represents the energy induced by centrifugal forces. In Ref. \cite{Li2018B1}, the authors do not take into account the influence of centrifugal forces, despite their considerable magnitude in comparison to the Coriolis forces. This approximation is valid only when considering the frequency $\omega=\omega_b$ instead of $\omega=0$ in the power spectral density since the centrifugal forces only affect the zero-frequency component.

Furthermore, using Eqs. \eqref{PXR}-\eqref{PYR} and the relation
\begin{eqnarray}
\begin{pmatrix}
\dot{x}\\
\dot{y}	
\end{pmatrix}
=\Omega
\begin{pmatrix}
-\sin\theta(t) & \cos\theta(t)\\
-\cos\theta(t) & -\sin \theta(t)	
\end{pmatrix}
\begin{pmatrix}
x_o\\
y_o	
\end{pmatrix}\nonumber\\
+\begin{pmatrix}
\cos\theta(t) & \sin\theta(t)\\
-\sin\theta(t) & \cos \theta(t)	
\end{pmatrix}
\begin{pmatrix}
\dot{x}_o\\
\dot{y}_o
\end{pmatrix}
\end{eqnarray}
one can easily check that the conjugate operators $x(y), p_{x(y)}$ also satisfy the basic commutation relation $[x,p_x]=[y,p_y]=i\hbar$. In addition, the Hamiltonian \eqref{HamiltonianR} can be examined by deriving its classical equations of motion using the Hamilton canonical equation, which has the same form as the equations of motion derived from the Heisenberg equation.

The Hamiltonian \eqref{HamiltonianR} can be rewritten as Eq. \eqref{HM} in terms of creation and annihilation operators with the transformation
\begin{eqnarray}
&&x=\sqrt{\frac{\hbar}{2m\omega_x}}(b_x+b_x^\dag),~~~~p_x=-i\sqrt{\frac{m\hbar\omega_x}{2}}(b_x-b_x^\dag),\nonumber\\
&&y=\sqrt{\frac{\hbar}{2m\omega_y}}(b_y+b_y^\dag),~~~~p_y=-i\sqrt{\frac{m\hbar\omega_y}{2}}(b_y-b_y^\dag),\nonumber
\end{eqnarray}
where the mechanical frequency is $\omega_{x(y)}=\sqrt{k_{x(y)}/m}$, and we set the equilibrium positions $x_o=y_o=0$ for brevity in deriving Eq. \eqref{HM}.

\providecommand{\noopsort}[1]{}\providecommand{\singleletter}[1]{#1}%

\end{document}